\documentclass[preprint,12pt,3p]{elsarticle}
\usepackage{lmodern}
\biboptions{sort&compress}

\usepackage[table]{xcolor}
\definecolor{myred}{rgb}{0.7961, 0.2627, 0.2039}  
\definecolor{mygreen}{rgb}{0.1294, 0.6039, 0.3294}  
\usepackage{booktabs}
\usepackage{amssymb}
\usepackage{amsmath}

\usepackage[colorlinks=true, citecolor=blue]{hyperref}   
\usepackage{cleveref}   

\usepackage[figurename=Figure, labelsep=space, font=scriptsize, labelfont=bf]{caption}

\usepackage{booktabs} 
\usepackage{textcase}

\usepackage{xcolor}

\begin{document}

\hypersetup{linkcolor=blue, urlcolor=blue,citecolor=black}

\begin{frontmatter}

\title{\Large{\textbf{Analog signal processing, storage, and demultiplexing via in-materia wave-based computing}}}
\makeatletter
\def\theaffn{\arabic{affn}}
\makeatother
\author[label1]{Saeed Behboodi}
\author[label1]{Mohamed Mousa}
\author[label1,label2]{Mostafa Nouh\texorpdfstring{\corref{mycorrespondingauthor}}{}}
\cortext[mycorrespondingauthor]{Corresponding author}
\ead{mnouh@buffalo.edu}

\address[label1]{Department of Mechanical and Aerospace Engineering, University at Buffalo (SUNY),\\ Buffalo, NY 14260-4400, USA}
\address[label2]{Dept. of Civil, Structural and Environmental Engineering, University at Buffalo (SUNY),\\ Buffalo, NY 14260-4300, USA}

\begin{abstract}
The ability of mechanical systems to perform basic computations has gained traction in recent years, providing a robust supplement to digital computing in off-grid, low-power, and extreme environments where electronics are inaccessible or risk failure. However, prior work has focused on Boolean logic applied via bistable structures, snap-through mechanisms, or origami lattices. Digital signal processing (DSP) underpins a wide range of computational tasks, providing the foundation for information encoding, manipulation, and analysis. In contrast to conventional DSP approaches, we present a new paradigm for analog information processing, where an elastic substrate houses a spatial wave-based platform for mechanical signal manipulation in the temporal domain. The proposed metasurface-based system exploits temperature-dependent phase transitions of shape memory alloys (SMAs) to realize key signal-processing functionalities, including gating, retention, and programmable delay of impinging mechanical loads. In doing so, we address a new set of challenges by manipulating transient mechanical waves through dispersive and lossy media. Building on this capability, we introduce the concept of mechanical wave-based memory, in which a locally resonant metamaterial's dispersion characteristics are actively reconfigured to achieve programmable delays while preserving spatial waveform integrity, and demonstrate its utility in the context of input shaping and frequency demultiplexing of superimposed waves. More broadly, the modular and scalable nature of the demonstrated framework paves the way for advanced mechanical computing architectures, enabling higher-order operations such as signal classification, parallel processing, and reservoir computing through the controlled sequencing and temporal alignment of mechanical signals.
\end{abstract}


\end{frontmatter}

\newpage

\section{Introduction}
\label{introduction}

The resurgence of mechanical computing over the past decade represents a paradigm shift in how we approach information processing in resource-constrained and extreme environments. Mechanical computing has evolved from historical devices, such as celestial gearboxes and tide-predicting machines, into a viable platform capable of processing information directly in the mechanical domain. Enabled by advances in metamaterial design and fabrication, contemporary mechanical systems can perform computation with low energy consumption, minimal latency, and without repeated transduction between mechanical and electrical signals \cite{yasuda2021mechanical,li2019information, goh2022nonlocal}. Hardware-level computing leverages the ability of mechanical systems to directly interact with the environment \cite{quevedo2019roadmap, caloz2013analog}, utilizing continuous quantities, such as acoustic pressure, vibrations, and forces, to interpret and manipulate information in their native physical form \cite{rajabalipanah2021analog, zangeneh2021analogue}. This integration with external stimuli enables seamless operation in environments where electronics are impractical, including remote, off-grid locations with scarce energy resources, or harsh conditions involving elevated temperatures or ionizing radiation \cite{mahon2019soft, silva2014performing}. 

Recent explorations of mechanical computing have yielded rapidly evolving computational capabilities across diverse architectures \cite{el2021digital, zhang2023mechanical}. Multiple research directions have realized logic gates and Boolean operations through bistable and multi-stable elements \cite{zhang2021hierarchical, chen2021reprogrammable, yasuda2020transition}, origami-inspired structures \cite{treml2018origami, melancon2021multistable, kamrava2017origami, liu2023discriminative}, and nonlinear dynamical systems \cite{lin2022anomalous, cotrufo2024reconfigurable}. In parallel, dynamic wave-based systems have emerged, utilizing propagating elastic waves as information carriers. By exploiting the rich physics of elastic wave propagation, including scattering, dispersion, beaming, and non-reciprocity \cite{adlakha2020frequency, moghaddaszadeh2021nonreciprocal, lv2021implementing}, elastic metamaterials can shape, control, and manipulate mechanical signals in a manner that extends far beyond static responses through carefully engineered dispersion characteristics and band structures \cite{shen2024rasterized,li2025forward,soliman2026non}. These capabilities have established the framework for wave-based mechanical computing (WBC), in which elastic waves encode information through their amplitude, phase, frequency, and arrival time \cite{sun2022artificial, tahmasebi2022parallel}. Within this domain, recent systems feature arithmetic problem-solving \cite{solli2015analog, mousa2024parallel}, neuromorphic inference \cite{mousa2025analog, wu2019neuromorphic, moghaddaszadeh2024mechanical}, and different implementations of logic gates \cite{fort2025thermoelastic}, all enabled by advances in phononic and adaptive metamaterials. 

Despite these advances, current WBC systems must overcome significant challenges on their path to reaching computing maturity. While effective at filtering, routing, spatially localizing, or temporally modulating incident signals \cite{thomes2022space,ye2025nonreciprocal}, they primarily manipulate information in space rather than in time, and generally lack the ability to actively control when information is stored, synchronized, or released. This inability to control computational timing represents a critical gap that fundamentally limits the overall complexity of operations WBC systems can execute. In signal processing and control theory, time delay, data retention, and temporal alignment are prerequisites for most sophisticated operations. In digital computing, memory and temporal control are essential for sequential logic and synchronization of parallel operations. Similarly, in mechanical computing, this limitation becomes evident in cascaded systems, where outputs from multiple computational elements must be precisely synchronized before interacting at a subsequent stage \cite{bilal2017bistable, jafari2026passive}. Consequently, programmable temporal control, including delay, storage, synchronization, and release, is a fundamental computational primitive rather than an auxiliary signal-processing function in mechanical computing systems \cite{yang2023analog, kwon2018nonlocal, yang2025metasurface}. 

Prior mechanical systems have achieved delays through dispersive spreading, resonant filtering, or rerouting \cite{han2023origami, gupta2026enduring}, but have nonetheless suffered from fundamental tradeoffs, including increased footprint and signal distortion. To address this gap, we propose a memory-integrated mechanical signal processing system that dynamically controls the flow of elastic waves through thermally activated metasurfaces, providing a computing platform in environments with access to fluctuating temperatures (e.g., jet engines, combustion exhausts, or waste heat). By exploiting crystallographic phase transitions of shape memory alloys (SMAs), the computing substrate can admit, trap, and release mechanical information in the form of waves. The building blocks of the metasurface-based WBC are carefully designed to provide zero or near-full transmission at different frequencies within their respective states, creating a dynamic gating mechanism that can admit incident signals, confine them for a controlled duration, and release them on command, while preserving their spatial and spectral structure. The reversibility and programmability of SMAs provide a robust mechanism which ensures the system operates repeatedly without degradation, making it suitable for a wide range of signal processing tasks. Because the system largely decouples temporal delay from propagation distance, the proposed architecture provides a compact, modular, and reconfigurable foundation for scalable wave-based mechanical computing.

The paper is organized into six sections. Following this introduction, we establish the fundamental principles governing mechanical signal delay in thermoelastic systems. We then exploit the switching capabilities of SMA-based unit cells to achieve programmable signal retention and on-demand release. Next, we apply this framework to two representative problems, demonstrating the system’s ability to perform spatiotemporal wave demultiplexing and tailor incident loads through input shaping. We subsequently integrate these capabilities into a three-dimensional configuration, illustrating the distillation of acousto-mechanical images into their constituent elements, providing a visually relatable computing exercise. Finally, we conclude with a summary of the main findings and their implications. The central contribution of this work lies in the engineered use of independently-controlled gates to enable programmable temporal control of elastic-wave information. Through numerical demonstrations, we establish controlled signal admission, retention, and release with a prescribed delay, achieved through SMA-based stiffness tuning and bandgap manipulation. This time-gating capability enables frequency-selective processing, wave demultiplexing, input shaping, and the selective retention and release of spatially encoded signals.

\section{Signal delay in mechanical systems}

In digital signal processing, time delay is a fundamental mechanism that plays a pivotal role in phase alignment, spectral shaping, and information management. This delay is usually applied by specific components such as delay lines \cite{palanisamy2024void}, buffers \cite{leib2017mechanical, zheng2021parameters}, or flip-flops \cite{li2024reprogrammable,mei2023memory}, resulting in precise control over the signal's timing. In addition to signal synchronization across different system components, this delay also serves as a powerful computational tool in applications as diverse as filtering, modulation, and sequence modeling. Similarly, time delay in wave-based mechanical computing is the physical retention of wave packets propagating through an elastic medium. In such wave-based systems, the ability to introduce a temporal interval between the incident and output signal effectively functions as short-term mechanical memory. In here, mechanical memory refers to the programmable, temporary, and controlled storage and subsequent release of information carried by propagating elastic waves, rather than persistent data storage in a static mechanical configuration such as multistable structures or origami lattices \cite{chen2021reprogrammable,li2024reprogrammable}. This sequence differs from a standard passive delay line, where output timing is determined primarily by the structural propagation path length \cite{palanisamy2024void}. On the other hand, the proposed system decouples temporal delay from propagation distance by trapping wave packets in an intermediate domain until an exit gate is thermally reopened. Furthermore, unlike digital memory, this mechanical storage is transient and finite, as storage duration is ultimately bounded by attenuation, leakage, and waveform distortion constraints. As will be shown throughout this effort, this temporal control is a prerequisite to advanced mechanical signal processors, enabling the synchronization and preprocessing of data prior to higher-level computations. However, realizing a truly programmable mechanical computing platform requires a mechanism to actively regulate signal retention in real time, representing a transition from static delay lines to dynamically tunable structures, as will be detailed next.

\subsection{Thermoelastic wave gates}
\label{design_gate}

\begin{figure*}
\centering
\includegraphics[width=\linewidth]{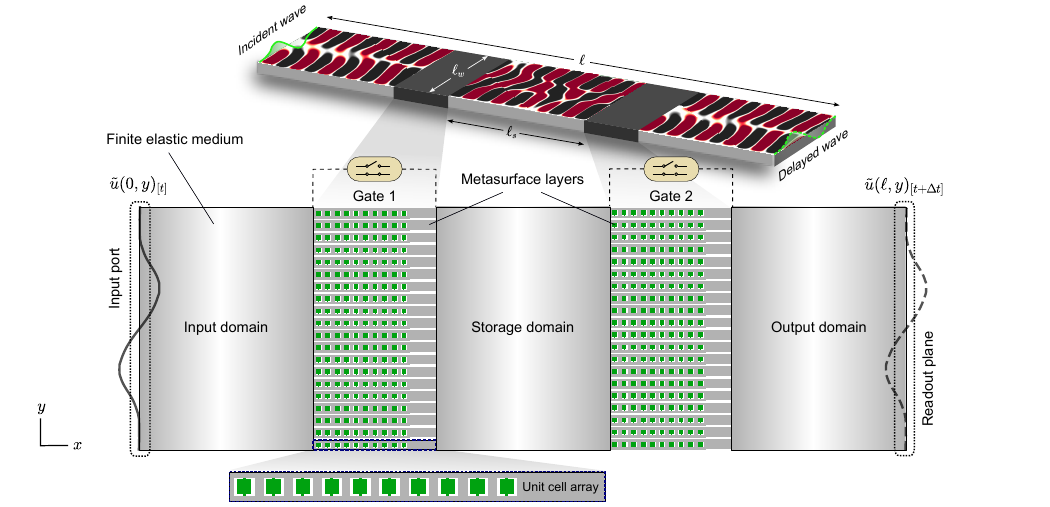}
\caption{\textbf{Mechanical signal delay via thermally programmable architected material.} Schematic of the underlying system of total length, $\ell$, consisting of input, storage, and output domains, embedded within a finite elastic medium. The storage domain has length, $\ell_s$, along $x$ and width, $\ell_w$, along $y$. The diagram depicts the propagation of a wave, $\tilde{u}(0,y)_{[t]}$, incident on the input port ($x=0$) and its release as a delayed signal, $\tilde{u}(\ell,y)_{[t+\Delta t]}$, at the readout port ($x=\ell$). Temporary information trapping is enabled via two metasurface layers (Gates $1$ and $2$), each comprising $20$ waveguides with $10$ locally resonant SMA unit cells per waveguide.} 

\label{Fig1}
\end{figure*}

Propagating waves within elastoacoustic structures function as data carriers. They transmit information by virtue of amplitude, phase, and frequency modulations in both spatial and temporal domains. As such, the ability to delay, store, and release information available in the waveform requires a means of wave trapping that preserves the wave's spatiotemporal integrity. The core concept we present encompasses a passive data storage domain flanked by two acoustic metasurface layers, forming a gate which admits or impedes the flow of information in the overarching system. The metasurfaces can be dynamically controlled in real time by triggering key elements in the structure, toggling the gates between transmission and blocking modes. The building blocks of the metasurfaces are elastic metamaterial cells which house shape memory alloy (SMA) resonators. Upon thermal activation, the metasurfaces undergo a crystallographic transition\cite{lunser2024transforming}, followed by a stiffness shift between the soft (martensite) and hard (austenite) phases \cite{candido2018tunable}, inducing a dispersion diagram shift which permits the unit cells to alternate between passband and bandgap states. This allows changes in ambient temperature to open and close these ``thermoelastic gates'', effectively admitting incident waves from an input to an intermediate (middle) region, where the information can be retained for a desired amount of time. Following another thermal change, an exit gate can be opened to release the stored waves, as illustrated in Fig.~\ref{Fig1}. This controlled delay turns the mechanical structure into a programmable information retention system, providing a basis for time-based mechanical signal processing. While this concept is fundamentally straightforward, the complex nature of guided wave scattering and the difficulty of retaining the information embedded in the impinging waves in their original form pose challenges for realizing such functionality, from unit-cell design to the overall architecture.

To realize the delay mechanism, a structure of length $\ell$ is configured to have three functional elastic compartments: an input domain, a storage domain of length $\ell_s$ and width $\ell_w$, and an output domain, connected via two metasurface layers (gates), as shown in Fig.~\ref{Fig1}. The left edge of the structure ($x=0$) marks the input port where external excitations are incident on the system, while the right edge ($x=\ell$) represents the readout port where computational outcomes are outputted. While the three aforementioned domains are physically continuous, they undertake distinct computational roles due to the presence of the dynamically controlled gates. The first metasurface (Gate $1$) is located at the entrance of the storage domain, with the second metasurface (Gate $2$) positioned at its exit. Together, they form a programmable path capable of temporarily trapping mechanical energy. The structure is initially excited at the input port with a distributed mechanical force whose amplitude varies along the $y$-axis, where $y$ represents the vertical coordinate along the input boundary. This leads vibrational waves to propagate along the longitudinal direction, $x$, causing material deformation, $u$. These deformations vary in space and time leading to a wavefield, where, at steady state, $u=\tilde{u}(x,y) e^{\mathrm{i} \omega t}$ represents the displacement at any given location for waves oscillating at a frequency $\omega$, and $\tilde{u}$ indicates the spatial component of the wavefield. As such, the vertical displacement ``profile'' at the input port, $\tilde{u}(0,y)$, represents an input signal to the system at some arbitrary time instant, $t$, as shown at the left boundary of the schematic in Fig.~\ref{Fig1}. In response, the system regulates the gates to either transmit the signal or temporarily store it in the designated storage domain for a desired duration, $\Delta t$. The preserved signal can be released into the output domain and picked up at the opposing end of the structure in the form of a vertical displacement profile at the readout port, i.e., $\tilde{u}(\ell,y)$, at time $t+\Delta t$, as depicted in the same figure.

\subsection{Unit cell design and thermal activation}
\label{design}

\begin{figure*}
\centering
\includegraphics[width=\linewidth]{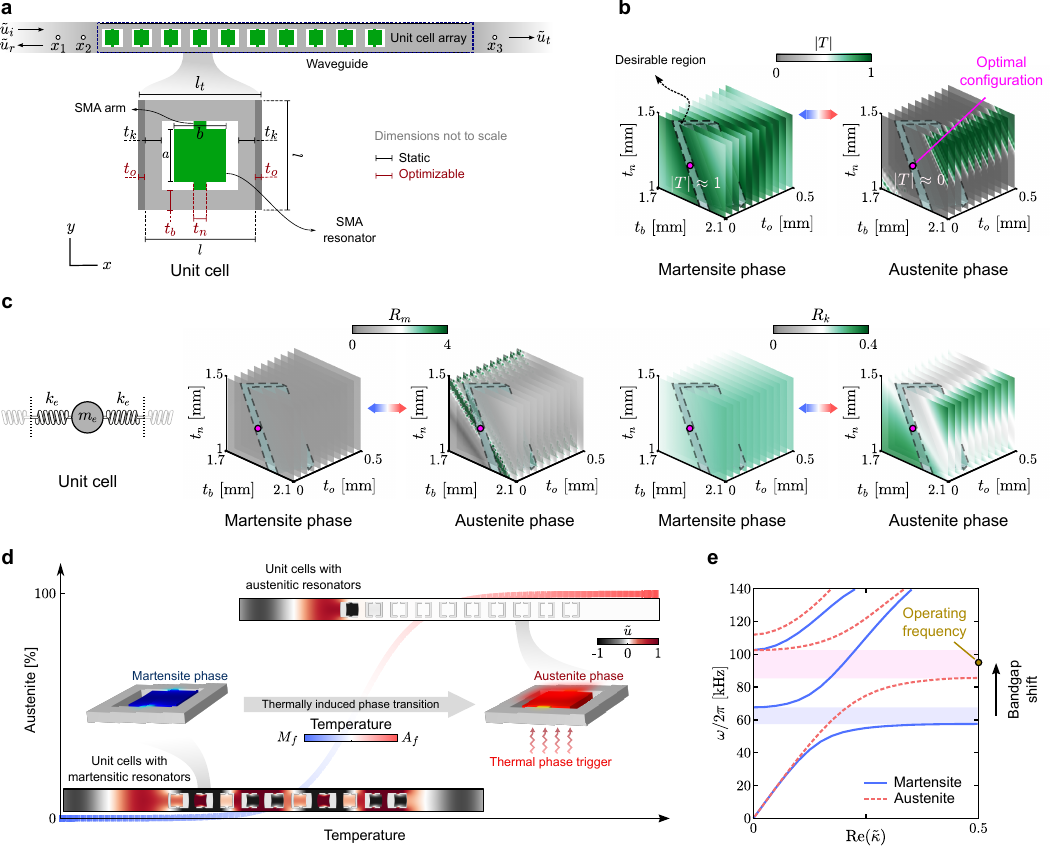}
\caption{\textbf{Unit cell geometry and phase-dependent wave transmission profiles. (a) Configuration of a single waveguide and its constituent unit cell. Each cell consists of a Nitinol resonator ($a=b=5$ mm), housed in a steel frame of outer width $l_t=l+2t_o$, with $l=10$ mm and $t_o=0.065$ mm. The frame is defined by the horizontal and vertical thicknesses, $t_k=1.51$ mm, and $t_b=1.9$ mm, respectively, and is connected to the resonator via two narrow necks of thickness $t_n=1.2$ mm. Probes at $x_1$, $x_2$, and $x_3$ evaluate the waveguide's transmission amplitude, $|{T}|$, as a function of the aforementioned geometric parameters for different material phases, as shown in (b). The ``Desirable region'' marks unit cell configurations with a passband state ($|{T}| \approx 1$) in the martensite phase and a bandgap state ($|{T}| \approx 0$) in the austenite phase at the target frequency of $\omega_t/{2\pi}=92$ kHz. (c) Equivalent mass-spring representation of the unit cell and corresponding maps of the normalized effective mass, $R_m$, and stiffness, $R_k$, ratios as functions of the unit-cell geometric parameters in both SMA phases. (d) Thermally driven transition of the unit cell resonator from a soft martensitic phase (blue) at a temperature $M_f$ to a stiff austenitic phase (red) at a temperature $A_f$, triggering the bandgap shift depicted in (e).}}
\label{Fig2}
\end{figure*}

Structurally, each metasurface layer comprises $20$ parallel waveguides spaced at intervals of $1$ mm. Each waveguide features $10$ periodic unit cells, collectively forming a subwavelength interface for accurate, high-resolution wavefront manipulation. The core component of the metasurface unit cells is a square-shaped thermally activated SMA resonator made of Nitinol (NiTi), housed within a steel matrix with an elastic modulus of $E_{st}=210$ GPa, a density of $\rho_{st}=7850$ kg/m$^3$, and a Poisson's ratio of $\nu_{st} = 0.28$ (Fig.~\ref{Fig2}a). NiTi, a widely used SMA composed of an approximately equiatomic mixture of nickel and titanium, is well known for its martensite-austenite transformation \cite{liang1990constitutive}. At a sufficiently low temperature, $M_f$, NiTi resides in the martensitic phase with a monoclinic lattice which has a relatively compliant structure, an elastic modulus of approximately $E_m=26.3$ GPa, and a density of $\rho_m=6400$ kg/m$^3$, allowing for significant shear deformation \cite{fort2025thermoelastic}. Conversely, as the temperature increases, the internal crystallographic structure fully transitions to a highly symmetric cubic lattice, namely the austenitic phase, once a temperature threshold, $A_f$, is exceeded. This threshold is approximately $45~^\circ$$\mathrm{C}$ for the composition considered here. The transition leads to notable stiffening, raising the elastic modulus to about $E_a=67$ GPa, while the density remains practically unchanged. The process is schematically represented in Fig.~\ref{Fig2}d.

The SMA unit cells are designed such that the dispersion bands of the martensite phase place a given target frequency ($\omega_t/{2\pi}=92$ kHz in this case) in a passband, whereas the transformation to austenite imposes a bandgap. Consequently, the same waveguide allows incident elastic waves to propagate without attenuation at low temperatures, while forbidding such propagation upon heating, providing a reversible mechanism for temporal control of information flow within the system, driven solely by thermal activation. To achieve this transmission contrast, a finite element (FE) model is developed in COMSOL Multiphysics\textsuperscript{\textregistered{}} using the Structural Mechanics module to characterize phase-dependent wave propagation. For a unit cell of length $l_t = l+2t_o$,
(see dimensions in Fig.~\ref{Fig2}a), the geometric design variables highlighted in red, including the neck thickness, $t_n$, and surrounding frame parameters, $t_b$ and $t_o$, are optimized to provide desired transmission properties, while the frame thickness, $t_k$, and the remaining dimensions are held constant. For a given wavenumber, $\kappa$, bandgap frequencies are determined from the dispersion relation via a Bloch-wave analysis, as shown in Fig.~\ref{Fig2}e, with $\text{Re}(\tilde{\kappa}) = \kappa l_t / 2\pi$ denoting the dimensionless wavenumber. This normalization fixes the Brillouin zone edge at $0.5$, enabling consistent evaluation of bandgaps as the unit cell length dynamically varies during the optimization process. The aforementioned parameters tune the phase-dependent dynamic response of the unit cell, which dictates the switching between high and low transmission states in the martensite and austenite phases, respectively. To maximize this transmission contrast, an intricate balance of these geometrical parameters is required. This is achieved through a frequency domain optimization procedure in which the Nelder-Mead algorithm is employed, due to the nonlinear dependence of transmission characteristics on unit cell geometry. The optimization is formulated as a multi-objective problem, and the parameter bounds are imposed to constrain the design space and prevent non-physical solutions.

The connection between individual waveguides and gate transmission can be understood in terms of metasurface dynamics by representing the unit cell with an equivalent mass-stiffness model \cite{moghaddaszadeh2024mechanical}. For each prescribed SMA phase, the unit cell is represented by an effective dynamic mass, $m_e$, and stiffness, $k_e$, obtained through transfer matrix equivalence. These quantities describe the collective axial response of the SMA resonator in the steel frame, rather than the resonator mass and stiffness alone. We define the normalized effective mass and stiffness ratios as follows,
\begin{equation}
R_m=\frac{m_e}{m_{\mathrm{ref}}},\quad \quad R_k=\frac{k_e}{2k_{\mathrm{ref}}},
\label{eq:RmRk}
\end{equation}
where $m_{\mathrm{ref}}={\rho }_{st}Al_{t}$ and $k_{\mathrm{ref}}=\frac{E_{s} A}{l_{t} (1-{\nu }_{s}^{2})}$ are the mass and axial stiffness, respectively, of a uniform reference steel waveguide of length $l_t$ and cross-sectional area $A$. These effective properties govern wave propagation within the waveguide through the dispersion relation,
\begin{equation}
2k_e \sin^2 \bigg(\frac{\kappa l_t}{2}\bigg) - \omega^2 m_e = 0.
\label{eq:dispersion}
\end{equation}

For a waveguide comprising $r$ identical cells ($r = 10$ in the present study), the complex displacement transmission coefficient can be expressed as,
\begin{equation}
T=\frac{2}{2\cos{(r{\kappa }_{e}l_{t})}+\mathrm{i}(\frac{z_{st}}{z_e}+\frac{z_e}{z_{st}})\sin{(r{\kappa }_{e}l_t)}},
\label{eq:T1}
\end{equation}
where $z_{st}=A\sqrt{\frac{{\rho }_{st}  E_{st}}{(1-{\nu }_{st}^{2})}}$ and $z_{e}=\frac{\sqrt{m_{e}k_{e}/2}}{\cos{({\kappa }_{e}l_{t}/2)}}$ are the host-waveguide impedance and the Bloch impedance of the equivalent waveguide, respectively. The analytical maps in Fig.~\ref{Fig2}c depict how these propagation and impedance characteristics combine to determine transmission at the operating frequency. The effective properties corresponding to the two SMA phases are marked on the maps, linking the phase-dependent unit-cell response to the gate’s switching behavior. From a unit-cell perspective, the thermally induced phase transition changes the SMA elastic modulus and resonator stiffness, thereby regulating the resonator's interaction with the surrounding steel frame. At the operating frequency, the equivalent normalized dynamic mass, $R_m$, increases from $0.29$ to $21.93$, whereas $R_k$ only changes slightly from $0.25$ to $0.26$. The transmission contrast is therefore associated primarily with a pronounced change in effective dynamic inertia and consequently, the impedance, $z_e$, of the unit cell, which governs the Bloch wavenumber and propagation characteristics, as formulated in Eq.~(\ref{eq:dispersion}). As a result, the operating frequency lies within a passband in the martensite phase and within a bandgap in the austenite phase. Equation~(\ref{eq:T1}) of this analysis confirms this observation, as this shift in the effective impedance, $z_e$, significantly lowers the transmission amplitude, which explains how the transition of the resonator phase produces the desired passband-to-bandgap transition.

Transmission coefficients associated with a unit cell design in FEM can be quantified using displacement information at three points, with two probes located upstream of the unit cell array at positions $x_1$ and $x_2$, and one probe downstream at position $x_3$. This configuration assumes the existence of incident ($\tilde{u}_i$), transmitted ($\tilde{u}_t$), and reflected ($\tilde{u}_r$) wave components. We note that, when testing a particular array configuration, the waveguide is padded with perfectly matched layers at the inlet and exit terminals to mitigate back reflections from the end terminal. As such, three points are sufficient to estimate the transmission coefficient amplitude for the waveguide from the following equation \cite{mousa2024parallel},
\begin{equation}
|{T}|=\bigg|\frac{\tilde{u}_t}{\tilde{u}_i}\bigg|=\Bigg|\frac{2\tilde{u}(x_3) e^{\mathrm{i} \kappa x_3} \sin\big(\kappa (x_1-x_2)\big)}{\tilde{u}(x_2)e^{\mathrm{i}  \kappa x_1}-\tilde{u}(x_1)e^{\mathrm{i} \kappa x_2}}\Bigg|,
\end{equation}
where $\tilde{u}(x_i)$ is the complex displacement at probe location $x_i$, with $i=1,2,$ and $3$, and $\mathrm{i}$ is the imaginary unit. The main goal of this optimization is to impose a near-zero transmission in the austenite phase and a near-unity transmission in the martensite phase, at the same targeted frequency ($92$ kHz). Once an optimal configuration is identified, a fine parametric sweep of the $t_n$, $t_b$, and $t_o$ parameters is carried out to study the local sensitivity and robustness of the optimized design. This sweep extends the search around the optimal point to determine how unit cell fabrication tolerances and small variations in design parameters affect the target contrast. As illustrated in Fig.~\ref{Fig2}b, the resulting design maps for both phases exhibit clear regions of strong contrast, where a martensite waveguide provides minimal attenuation ($|{T}_m| \approx 1$), as opposed to an austenite waveguide exhibiting near-zero transmission ($|{T}_a| \approx 0$) and, subsequently, a complete block of incident waves. This approach forms the basis for unit cell design in all the subsequent sections. 

\section{Programmable time delay and signal retention}
\label{concept}

\begin{figure*}
\centering
\includegraphics[width=\linewidth]{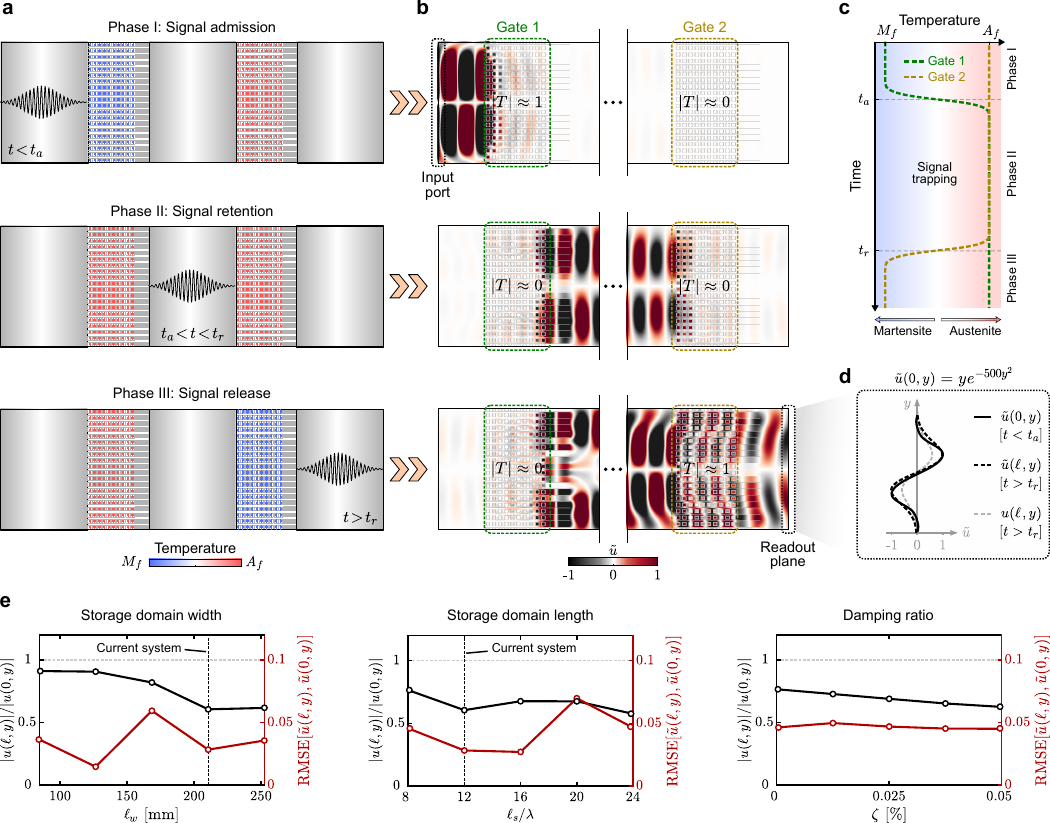}
\caption{\textbf{Programmable signal delay and retention in an elastic substrate.} (a) Three distinct stages of the wave delay mechanism, (b) corresponding wavefields, and (c) time history of the thermal environment driving phase transitions of unit-cell resonators in both gates. $t_a$ and $t_r$ denote admission and release time stamps, respectively. (d) Comparison of the normalized incident (solid) and readout (dashed) displacement profiles, $\tilde{u}(0,y)$ and $\tilde{u}(\ell,y)$, confirming near-full signal preservation and the effectiveness of the delay mechanism. The non-normalized readout, ${u}(\ell,y)$, is also plotted for reference (faded, dashed) showing signal attenuation. (e) Parametric sensitivity analysis evaluating the impact of structural damping ratio, $\zeta$, and storage domain dimensions, $\ell_w$ and $\ell_s$, on the achieved delay signal strength, $|u(\ell,y)|/|u(0,y)|$, and profile, $\mathrm{RMSE}[\tilde{u}(\ell,y),\tilde{u}(0,y)]$.}
\label{Fig3}
\end{figure*}

Having established the fundamental switching capabilities of the SMA-based unit cells, the ability of the aforementioned dual-gate system to force a programmable delay of an incident signal is evaluated. For this configuration, we pursue two specific objectives: First, to demonstrate the effective confinement of an elastic wave within the designated ``storage domain'', and second, to implement a controlled release of this trapped signal after a prescribed time period, while preserving the spatial integrity of the waveform throughout the process. To initialize our testing scheme (Phase I: Signal admission), the system is subjected to a harmonic excitation at the target frequency. Rather than exciting the input domain with a simple plane wave, an input displacement profile $\tilde{u}(0,y) = y e^{-500 y^{2}}$ is introduced at the input port, to assess the system's ability to preserve spatiotemporal information upon wave scattering. With the first metasurface layer initially maintained at an ambient temperature, $M_f$, the incident $\tilde{u}(0,y)$ wave propagates unimpeded through gate $1$ (where the resonators exhibit a passband) and is gradually admitted into the storage domain at a time stamp $t_a$ (Fig.~\ref{Fig3}a). The second layer resides in a thermally elevated environment, $A_f$, effectively blocking the waves' exit from the storage domain via a bandgap exhibited at the same target frequency. As the signal enters the region flanked by both gates, as depicted in Fig.~\ref{Fig3}b, the first metasurface layer transitions to an austenitic phase via thermal stimulation (Fig.~\ref{Fig3}c), putting the system in a trapping state. During this process (Phase II: Signal retention), the two gates act as reflective interfaces, effectively isolating the storage domain from the system's input and output in the form of standing waves. The signal retention period spans a prescribed duration $t_a < t < t_r$, during which, the near-zero forward and backward transmission exhibited by both gates ($|{T}_a| \approx 0$) shields the preserved information from ambient noise and external factors detrimental to signal integrity. While the thermally-triggered storage mechanism highlights the short-term memory capability, it also illustrates a breakage of dependency between signal delay (temporal control) and propagation length, allowing variable time delays to be independently achieved without a larger structural footprint.

After a programmed delay, the wave can be released at time stamp $t_r$ into the output domain, marking the last phase (Phase III: Signal release). As captured in Fig.~\ref{Fig3}c, Gate $2$ opens up upon thermal transition, discharging the stored waves towards the readout port. The effectiveness of the delay is numerically verified by comparing the incident and readout displacement profiles by time-averaging the waves at these two locations over comparable time durations before admission ($t<t_a$) and post-release ($t>t_r$). The spatial envelope of the released wave $\tilde{u}(\ell,y)$ matches that at the incident port, as shown in Fig.~\ref{Fig3}d, albeit with some inevitable damping losses. We note that since our focus is on qualitative features of the transmitted information, the $\tilde{u}$ profiles shown in the plots are always normalized with respect to their maximum value to provide an easily readable $[-1,1]$ scale. Supplementary Movie~$1$ provides an animation of the time-varying displacement field during signal admission, retention, and release, illustrating the programmable delay mechanism shown in Fig.~\ref{Fig3}.

While the current model assumes undamped elastic wave propagation, dissipation is a critical physical parameter that inevitably affects system performance. To address this, we have conducted a parametric sensitivity analysis evaluating the impact of structural damping ratio, $\zeta$, as well as layer dimensions (storage width and length), on the achieved delay in terms of transmitted amplitude and waveform distortion. In this sensitivity study, damping is modeled using COMSOL's time-domain maximum loss factor formulation at the operating frequency. For each damping level and geometrical configuration, the system's ability to retain the input amplitude is evaluated using the peak amplitude transmission ratio, $|u(\ell,y)|/|u(0,y)|$, calculated from non-normalized incident and released signals. The spatial waveform distortion is quantified by calculating the root-mean-square error of the temporally normalized incident and released waveforms before and after the delay, $\mathrm{RMSE}[\tilde{u}(\ell,y),\tilde{u}(0,y)]$. As shown in Fig.~\ref{Fig3}e, the proposed system is capable of preserving the profile for different conditions; however, the choice of the geometry and damping ratio involves a trade-off between the achievable delay and released signal amplitude.

\begin{figure*}
\centering
\includegraphics[width=\linewidth]{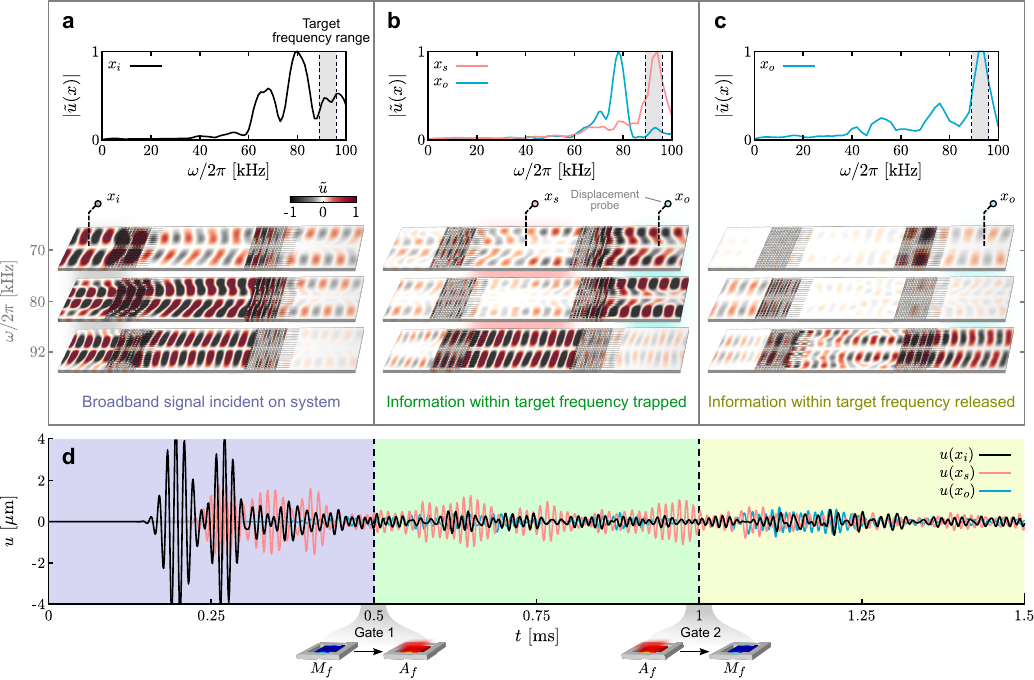}
\caption{\textbf{Frequency-selective information extraction.} The evolution of the incident signal from (a) $0$-$0.5$ ms, (b) $0.5$-$1$ ms, and (c) $1$-$1.5$ ms encapsulates the mechanically-encoded trapping, storage, and information extraction sequence within the SMA-based system. Throughout the process, wavefields are displayed at three distinct frequencies: $70$, $80$, and $92$ kHz, and show the stages from: (a) incidence to gradual signal admission, (b) signal trapping at the target frequency range and flushing out of non-targeted information, and (c) extraction and release of desired information. (d) Raw displacement amplitude at three probes placed at $x_i$, $x_s$, and $x_o$ within the input, storage, and output domains, respectively, providing a visualization of the same sequence in the time domain.}
\label{Fig4}
\end{figure*}

To further assess the performance of the architected system, we conduct an investigation to examine its spectral selectivity, i.e., its ability to store, delay, and release information contained in select frequency bands within a broader amalgamation of incident data. Unlike the single harmonic excitation adopted earlier, the system is again driven at the input port ($x=0$) with a broadband Gaussian-modulated wave packet, containing frequencies spanning the $60$-$100$ kHz range. The packet's dominant energy is centered around $80$ kHz as shown by the spectral amplitude plot in the top panel of Fig.~\ref{Fig4}a. Consistent with the previous task, we set a target frequency range at and around $\omega_t/{2\pi} = 92$ kHz, deeming information within this band desirable. The temporal profile of the incident signal is constructed via a superposition of delayed pulses, expressed as: $u = \tilde{u}(0,y) \sum_{n=1}^{2} e^{-t^2/2\sigma^2} \sin(\omega_c t)$, where $\tilde{u}(0,y)$ is the same input displacement profile defined earlier, $\sigma$ is the temporal envelope width, and $\omega_c$ is the central carrier frequency. As shown in Fig.~\ref{Fig2}e, the martensitic unit cell exhibits a passband ranging from $67.7$ kHz to $102.5$ kHz, whereas the austenitic unit cell exhibits a bandgap at the target frequency. 

%

Figure~\ref{Fig4} illustrates the entire sequence of the mechanically-encoded trapping, storage, and information extraction, within the SMA-based system. The spatiotemporal evolution of the incident signal is analyzed across the mechanical system at three distinct frequencies: $70$, $80$, and $92$ kHz, as displayed by the upper, middle, and lower wavefields in Fig.~\ref{Fig4}, respectively. The operation is broken down into three time periods representing the different stages of the process. Figure~\ref{Fig4}a shows the spectral content and wavefields corresponding to the initial incidence of the broadband signal onto the structure. The frequency response function $|\tilde{u}(x)|$ is evaluated at a displacement probe located at $x=x_i$ within the input domain, as indicated in the figure. The wavefields are obtained by applying a Fast Fourier Transform (FFT) on the time-domain displacement over the interval from $t = 0$ to $t=0.5$ ms. During this time, gate $1$ remains open to all incident frequencies, with all of its resonators residing in the martensite phase. As a result, the initial stage witnesses a near-full admission of all the incident data into the storage domain. The leftmost partition of Fig.~\ref{Fig4}d shows the raw (non-normalized) time-domain displacement corresponding to the same time interval at $x_i$ along with two additional probes located at $x_s$ and $x_o$ in the storage and output domains, respectively. The strong presence of waves at both $x_i$ and $x_s$, compared to $x_o$, confirms successful transmission of the waves through gate $1$.

Gate $2$ initially defaults to austenitic resonators, thus preventing information within the target frequency of $92$ kHz from further propagation downstream. Additionally, at $t = 0.5$ ms, gate $1$ transitions to austenite upon thermal stimulation, thereby achieving two-way blocking of the $92$-kHz component. As observed in Fig.~\ref{Fig4}b, wavefields corresponding to non-target frequencies ($70$ and $80$ kHz) continue to propagate through the system and are strongly present in the output domain, while the $92$-kHz component remains trapped within the storage domain, preserving its spatial variation. This is further confirmed by the frequency response functions $|\tilde{u}(x)|$ evaluated at $x_s$ and $x_o$ which show a prominent peak and negligible amplitude around $92$ kHz, respectively, reflecting the lack of this particular information subset in the output domain. During the interval between $t=0.5$ and $t=1$ ms, the system continues to ``flush out'' all the information contained within non-target frequencies, enabling a clean localization of the target information within the storage domain. The middle partition of Fig.~\ref{Fig4}d shows the robust displacement amplitude at $x_s$ throughout this process. 

The sequence concludes with the release of the stored information at a prescribed instant. Starting at $t = 1$ ms, as shown in Fig.~\ref{Fig4}c and the rightmost partition of Fig.~\ref{Fig4}d, as the resonators of gate $2$ revert to the martensite phase (upon removal of thermal stimulus), gate $2$ recalls a passband state at $92$ kHz and opens up, forcing the target frequency signal to flow to the readout port within the output domain. Spectral analysis at the downstream probe, $x_o$, averaged over the interval from $t = 1$ to $t = 1.5$ ms, confirms the successful release of the $92$-kHz component as evidenced by a pronounced peak at this target frequency. The process illustrates the ability of the system to function as a physical filter with short-term memory and signal retention capabilities. This allows a mechanical substrate to cleanly extract and preserve desired data housed within a broader information spectrum, providing an essential prerequisite to advanced signal processing analogous to digital computing platforms.  

The simulations employ a two-dimensional plane-stress formulation with a $1$ mm thickness and are solved linearly using the MUMPS solver. The domains are modeled as isotropic, linearly elastic materials. The architecture sides are assumed to be rigid with constrained displacement (i.e., $u_x = u_y = 0$), while low-reflecting boundary conditions are applied to the input and output terminals to reduce artificial reflections. The SMA switching is represented and applied to the system through tuning the SMA resonators' material properties at each state (martensite or austenite phase) at the prescribed gate-switching times. The simulation does not solve the heat transfer problem or predict the kinetics of the phase transformation, but rather applies the expected phase transformation effect on the resonator material properties at pre-configured switching instances. The computational domain is discretized using a mapped mesh with a maximum element size ranging of $1$ mm. This resolution is determined through a mesh independence study to optimally balance computational efficiency with numerical accuracy; refining the mesh beyond these parameters yielded negligible variations in the simulated response. For the transient solver, the maximum and initial time steps are set to $2\pi/(20\omega_t)$ and $2\pi/(50\omega_t)$, respectively, where $\omega_t/{2\pi}=92$ kHz.

\section{Delay-enabled functionality}
\label{application}

The framework realized in the previous two sections sets up the mechanical system for advanced signal processing with increasingly complex features. The ability to store and release wave-encoded information with high fidelity equips the metasurface-based system with a host of tools. To illustrate such versatility, we pose two representative problems in the domain of programmable spatiotemporal signal modulation. In the first example, we devise and test out a thermally-driven signal demultiplexer, with the aim of breaking down a combined transmission signal into its original data streams, through a frequency distillation process. In other words, we task the medium with reconstructing the individual (constitutive) frequency components of the aggregate signal and reading them out sequentially at prescribed time intervals. In the second example, we utilize the programmable delay mechanism to realize mechanical input shaping through constructive and destructive interference between delayed wave signals. In the next two sections, we demonstrate both applications and illustrate why the system is uniquely positioned to tackle similarly formulated challenges. 

\subsection{Spatiotemporal wave demultiplexing}
\label{demultiplexing}

\begin{figure*}
\centering
\includegraphics[width=\linewidth]{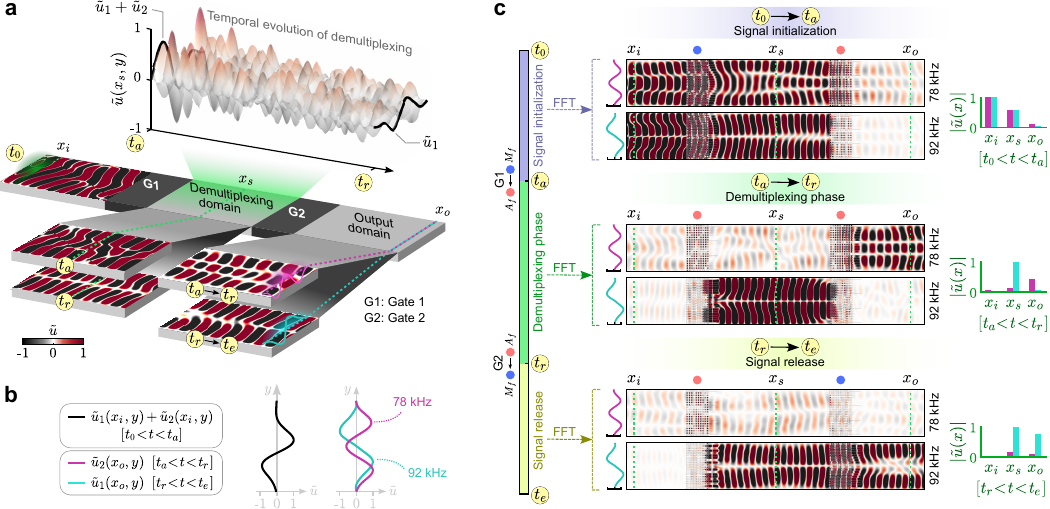}
\caption{\textbf{Mechanical spatiotemporal wave demultiplexing.} (a) Wavefield at critical milestones starting from $t_0$ (signal initialization) to $t_e$ (exit), passing through $t_a$ (admission) and $t_r$ (release). The close-up inset shows the temporal evolution of the demultiplexing of the incident signal, $\tilde{u}_1+\tilde{u}_2$, into $\tilde{u}_1$ along the indicated longitudinal cutline (green dashed line) over the interval $[t_a, t_r]$. (b) Spatial $y$-profiles of the incident and output signals at relevant time instants. (c) Sequential spectral and spatial analysis of the individual wave components associated with $\tilde{u}_1$ ($92$ kHz) and $\tilde{u}_2$ ($78$ kHz) obtained via FFT for the initialization, demultiplexing, and release phases of the process. Normalized amplitudes (rightmost bar charts) at the defined cutlines (green dashed lines) confirm the separation of the signal components.}
\label{Fig5}
\end{figure*}

Signal demultiplexing is vital because it permits large amounts of data to share the same transmission medium such as physical cables, or in this case, a mechanical or acoustic system. The core idea is to physically decompose a combined input wave and return it to its original components. Such capability drastically reduces infrastructure costs by allowing complex signal combinations to travel down a single transmission path and be separated at the destination. As demonstrated in the previous section, it is also particularly relevant in scenarios where the desired input signal is accompanied by undesired external loads or excitations originating from the surrounding environment, providing an ability to isolate important information from unwanted noise. As shown in Fig.~\ref{Fig5}a, the system established earlier is subjected to an incident wave at an initial time instant, $t_0$. The input wave is a combined signal composed of two frequency bands, a $92$-kHz component, $\tilde{u}_1$, and a $78$-kHz component, $\tilde{u}_2$. The two components have the same amplitude and distinct spatial profiles along the $y$-axis, and the aggregate signal at a displacement probe, $x_i$, placed within the input domain, is shown in Fig.~\ref{Fig5}b. We note that the results shown here readily extend to input signals with more than two bands by employing several bandgaps, which are typically exhibited by elastic metamaterials, but we focus here on a bi-frequency signal for ease of visualization. As depicted in Fig.~\ref{Fig2}d, the unit cells within both gates exhibit a passband in the low-temperature martensitic phase, $M_f$, at both frequencies, and a bandgap corresponding to $92$ kHz only in the high-temperature austenitic phase, $A_f$. The configuration implies an ``unconditional'' passband state at $78$ kHz giving $\tilde{u}_2$ permission to propagate through the different layers throughout the duration of the study regardless of ambient thermal conditions.

The demultiplexing process is governed by a temporal sequence of thermal stimuli, dividing the wave propagation scheme into three distinct stages, spanning the total interval from $t_0$ to $t_e$ (ending time), as illustrated in the timeline of Fig.~\ref{Fig5}c. At $t_0$, with the resonators pre-configured to $M_f$ and $A_f$ in the first and second gates, respectively, the incident signal is admitted in its combined form through gate $1$. The window from $t_0$ to $t_a$ represents the admission time interval, with $t_a$ approximately marking the time at which the combined signal travels downstream to reach the end of the region labeled as the demultiplexing domain. The upper panel of Fig.~\ref{Fig5}c shows two separate wavefields obtained by applying an FFT to the time-domain data over the $t_0 < t < t_a$ interval and evaluating the wavefields at $78$ kHz (top) and $92$ kHz (bottom). This visualization reveals that both components within the bi-frequency input co-exist in the demultiplexing domain at a near-identical amplitude, as confirmed by the bar charts alongside the wavefields. 

\newpage
Upon reaching $t_a$, the thermal status of gate $1$ is changed as indicated on the timeline of Fig.~\ref{Fig5}c, marking the onset of the demultiplexing process over the time period from $t_a$ through $t_r$. Due to the unconditional passband at $78$ kHz, this change in gate $1$ aligns its bandgap state with gate $2$, practically shutting them both off for the $92$-kHz component. As a result, during this window, $\tilde{u}_2$ proceeds to exit while $\tilde{u}_1$ remains confined to the demultiplexing domain. The middle panel of Fig.~\ref{Fig5}c shows the clear emergence of the $\tilde{u}_2$ component at the readout port within the output domain. In tandem, displacement probes at $x_s$ and $x_o$, located within the demultiplexing and output domains, respectively, confirm the same behavior with the bar charts showing a growth of $\tilde{u}_2$ at $x_o$ and a clear suppression of $\tilde{u}_2$ at $x_s$. The inset in Fig.~\ref{Fig5}a, measured at $x_s$, depicts the temporal evolution of the signal from the combined form at $t_a$ to a residual single-frequency component at $t_r$, marking the completion of the signal transformation. Finally, the process concludes with the release of the trapped $92$-kHz component, $\tilde{u}_1$, during the interval from $t_r$ to $t_e$, as a result of the change in the thermal status of gate $2$ as shown in the timeline of Fig.~\ref{Fig5}c.

\subsection{Mechanical input shaping}
\label{shaping}

Input shaping is an open-loop control technique used to reduce vibrations in computer-controlled environments through a set of designed commands. The approach works by shaping the input signal so that it cancels itself out, via a series of convolved impulses whose amplitude and time stamps are carefully computed \cite{al2018band}. Here, we present a second application of programmable signal delay in mechanical systems, allowing a thermally-responsive elastic medium to dispose of its own undesirable vibrations through ``mechanical'' input shaping. Such implementation critically renders the system self-sufficient in terms of its damping ability, since the system in principle exploits a latter subset of an impinging load to cancel vibrations excited by the early portion of the same signal, illustrating an important capability enabled by smart mechanical computing. 

\begin{figure*}[ht!]
\centering
\includegraphics[width=\linewidth]{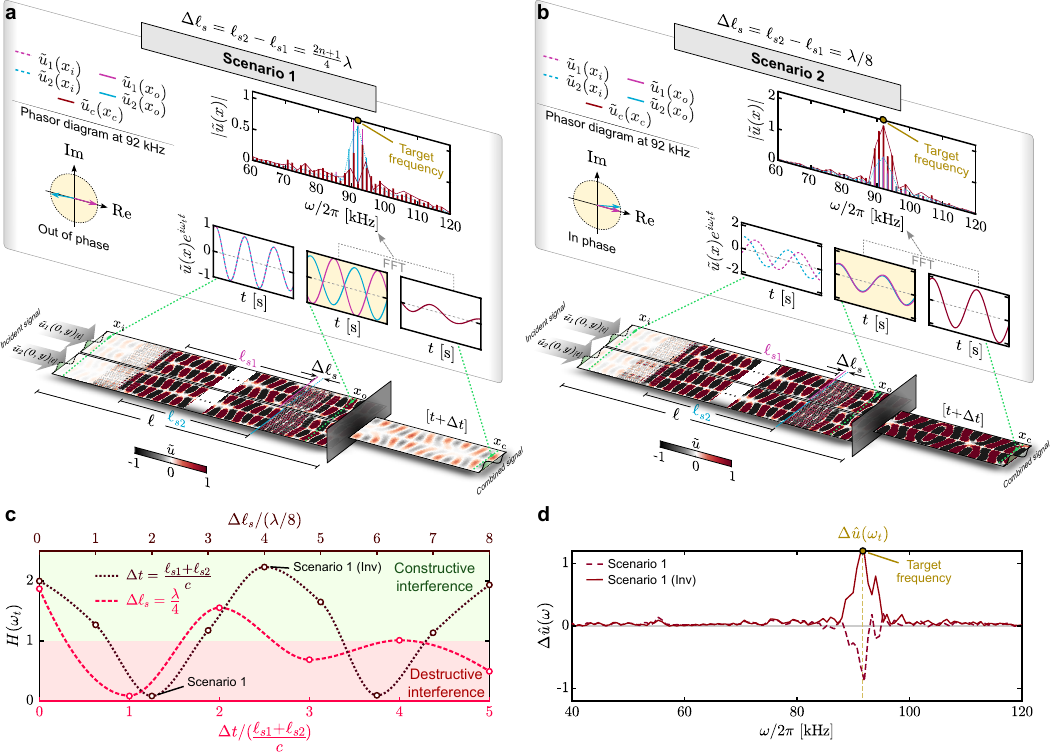}
\caption{\textbf{Signal cancellation and realignment via mechanical input shaping.} Relative phase control between two waves is achieved using parallel channels with different storage-domain lengths, $\ell_{s1} \neq \ell_{s2}$, and an engineered time delay, $\Delta t \approx \frac{\ell_{s1}+\ell_{s2}}{c}$, prior to signal combination. (a) Scenario $1$: Two identical in-phase waves, $\tilde{u}_1(0,y) = \tilde{u}_2(0,y)$, propagate through two channels with a tuned path-length difference of $\Delta \ell_{s} =\frac{2n+1}{4}\lambda$, effectively creating a $\pi$ phase shift and resulting in destructive interference. This is shown by the spatial wavefield (bottom), temporal snapshots of the signal at $x_i$, $x_c$, and $x_o$ (middle), and the FFT spectrum (top), comparing the individual channel components with the attenuated aggregate signal. (b) Scenario $2$: Two identical waves, $\tilde{u}_1(0,y) = \tilde{u}_2(0,y)$, with a $\pi/2$ phase difference, propagate through two channels with a tuned path-length difference of $\Delta \ell_{s} =\lambda/8$, effectively realigning the two channels and creating constructive interference. The resulting amplification is depicted by the spatial wavefield, temporal snapshots, and the FFT spectrum. (c) Dependence of the transfer function, $H(\omega_t)$, on both the path-length difference and time delay, illustrating the controllable transition between constructive and destructive interference, and the ability to achieve different input shaping outcomes via either mechanism. (d) Evaluation of the frequency-dependent normalized amplitude change, $\Delta\hat{u}(\omega)$, of the input shaping action corresponding to Scenario $1$ and its inverse, Scenario $1$ (Inv), highlighting suppression and amplification at the target frequency, $\omega_t$.}
\label{Fig6}
\end{figure*}

A theoretical basis of input shaping is the superposition of appropriately weighted and delayed copies of an input. A shaper is represented by an impulse sequence,
\begin{equation}
s(t)=\sum_{j=1}^{p} A_{j}~\delta (t-t_{j})
\end{equation}
where $A_j$ and $t_j$ denote the amplitude and time location of the $j$th impulse, respectively, and $\delta(\bullet)$ denotes the Dirac delta function. By applying the shaper to an original input, $u(t)$, the shaped input is,
\begin{equation}
u_s(t)=(s\ast u)(t)=\sum_{j=1}^{p} A_{j}u(t-t_{j}),
\end{equation}
where $*$ denotes convolution \cite{al2018band,singer1990shaping}. The corresponding frequency-domain shaping function is therefore given by,
\begin{equation}
S(\omega)=\sum_{j=1}^{p} A_je^{-\mathrm{i} \omega t_{j}}.
\end{equation}

For instance, a two-impulse shaper with equal weights, $A_1=A_2$, and a delay of $2\pi/(2\omega_t)$ yields $S(\omega_t)=0$ in an undamped system since the responses associated with the two delayed components cancel out at the target frequency. We employ this concept to manipulate mechanically-inputted waves to a structure, resulting in selective amplification or suppression of the input excitation at a targeted frequency. In this section, two representative mechanical input shapers are demonstrated. In the first scenario, a harmonic excitation of an undamped linear system is suppressed by delaying the propagating wave by a half-period and superimposing it on the original input, producing destructive interference of two out-of-phase signals. The process begins with an impinging displacement applied at an arbitrary time instant, $t$, with a $y$-profile that matches the one used in Sec.~\ref{concept}. The mechanical computing substrate consists of two structures which are subjected to the same input, resulting in identical initial displacement at the input ports of both, i.e., $\tilde{u}_1(0,y)_{[t]}=\tilde{u}_2(0,y)_{[t]}$. Each of these parallel channels imposes a controlled time delay, $\Delta t \approx \frac{\ell_{s1}+\ell_{s2}}{c}$, in which $c$ is the in-plane longitudinal wave speed of the thin steel plate, calculated from $c = \sqrt{E_{st} / [\rho_{st}(1-\nu_{st}^2)]}$, where $E_{st}$, $\rho_{st}$, and $\nu_{st}$ take the same values defined in Sec.~\ref{design}. As shown in Fig.~\ref{Fig6}a, the channels are set to have the same total physical length, $\ell$, but the lengths of their storage domains, $\ell_{s1}$ and $\ell_{s2}$, are different, introducing an engineered phase shift between the two propagating signals. The waves then recombine at the edge of both channels, yielding an output, $\tilde{u}_1(x_o,y)+\tilde{u}_2(x_o,y)$, which is equal to the superposition of the delayed signals. The combined signal, $\tilde{u}_c(x_c,y)$, can be measured at a readout location, $x_c$, in the non-computing portion of the medium, downstream of the input shaping channels. To accomplish this, the mechanical input shaper is tuned such that the difference in the storage domain lengths, $\Delta \ell_s=\ell_{s2}-\ell_{s1}$, is equivalent to a quarter-integer multiple of the wavelength corresponding to the target frequency, i.e., $(2n + 1)\lambda/4$, where $n$ is an integer and $\lambda$ is the pertinent wavelength. The length difference creates a relative phase shift of approximately $\pi$ between the signals propagating through the two channels. This can be explained by tracing the distances traveled by the signals to the output terminal. For a time delay of $\Delta t \approx \frac{\ell_{s1}+\ell_{s2}}{c}$, the first signal, $\tilde{u}_1(x_o,y)$, propagates through the storage domain back and forth, covering a distance of $3 \ell_{s1}$, in addition to $\Delta \ell_s$ immediately after being released. On the other hand, the second signal, $\tilde{u}_2(x_o,y)$, covers a distance of $3 (\ell_{s1}+\Delta \ell_s)$ while propagating within its storage domain. Therefore, the length difference, $\Delta \ell_s$, is effectively traversed three times by one signal and only once by the other. As a result, the difference in the distances covered by the two signals is $2 \Delta \ell_s$, or $\lambda/2$ for $n=0$, corresponding to the aforementioned $\pi$ difference. The out-of-phase condition causes the peak of one signal to align with the valley of the other, inducing near-perfect cancellation of the aggregate output wave at $x_c$, as shown in the figure insets.

The second example demonstrates a programmable signal alignment of two non-synced inputs. In this scenario, the substrate is subjected to two waves which, once again, share the exact $y$-profile but are incident on the system with an initial phase difference of $\pi/2$. This is synonymous with signal delays which often take place in individual channels within a larger network of transmission lines, due to impedance mismatches, trace differences, and other localized conditions, causing unintended disruptions due to asynchronous arrival times at signal combination ports. In this configuration, the storage-domain length differential, $\Delta \ell_s$, is set to a value equivalent to one-eighth of the wavelength, i.e., $\lambda/8$. This leads to an additional phase delay that compensates for the initial phase mismatch. Consequently, introducing a relative phase shift of nearly $\pi/2$ causes the propagating waves to sync up, achieving an in-phase condition in which the peaks and valleys of the two signals align. This results in strong constructive interference that nearly doubles the amplitude, as illustrated in Fig.~\ref{Fig6}b.

To bridge mechanical input shaping with classical input shaping theory, we can frame the behavior in terms of the relative delay between the two channel signals. Let $\tau_1$ and $\tau_2$ denote the respective two storage delays, and define the differential delay as $\Delta\tau = \tau_2- \tau_1$. For the channels with equal output amplitudes and an initial phase difference of $\phi_0$, the ideal combined response relative to either individual channel is expressed as,
\begin{equation}
G(\omega) =\big|1 + e^{\mathrm{i}(\phi_0 - \omega\Delta\tau)}\big| = 2\big|\cos\big(\frac{\phi_0 - \omega\Delta\tau}{2}\big)\big|.
\end{equation}

Destructive interference occurs when the relative phase is an odd multiple of $\pi$, whereas constructive interference occurs when it is an integer multiple of $2\pi$. For the propagation sequence described above, with $\phi_0=0$, the path length difference is $2\Delta\ell_s$, which corresponds to a time delay of $\Delta \tau \approx \frac{2\Delta {\mathrm{\ell }}_{s}}{c}$. As such, the relative phase will be,
\begin{equation}
\phi_{\mathrm{rel}}(\omega_t) = \phi_0 - \omega_t \Delta \tau.
\end{equation}

Accordingly, setting $\Delta\ell_s=\lambda/4$ produces destructive interference for initially in-phase inputs, while $\Delta\ell_s=\lambda/2$  restores constructive interference. In the second scenario, $\Delta\ell_s=\lambda/8$ provides a relative phase delay of $\pi/2$, thereby compensating for an initial phase lead of $\phi_0=\pi/2$ in the longer-delay channel. The system's tunability is manifested by mapping the magnitude of the transfer function,
\begin{equation}
H(\omega_t)=\frac{\mathrm{max}(|\mathcal{F} \{\tilde{u}_c(x_c,y)\}|)}{\mathrm{max}(|\mathcal{F} \{\tilde{u}_1(x_o,y)\}|)},
\label{eq:TF}
\end{equation}
at the target frequency, where $\mathcal{F}(\bullet)$ denotes the Fast Fourier Transform of the associated argument. In Fig.~\ref{Fig6}c, $H(\omega_t)$ is evaluated as a function of both the normalized path-length difference (dotted) and the corresponding relative time delay (dashed). The resulting curves highlight a continuous and highly-controllable transition between two ends of the spectrum, namely, ideal destructive ($H(\omega_t) \approx 0$) and ideal constructive ($H(\omega_t)\geqslant2$) interferences. The figure also illustrates the different possible paths to the same desired interference pattern of the mechanical input shaper by tracing the sought $H(\omega_t)$ value to a given metasurface layer length differential, $\Delta \ell_s$, or time delay, $\Delta t$. This provides a tool by which different functionalities can be obtained from the same substrate by adjusting the temporal delay via thermal stimulation, rather than a physical reconstruction of the entire substrate to meet different outcomes.

The example demonstrated in Fig.~\ref{Fig6}a can also be understood in light of the transfer function given by Eq.~(\ref{eq:TF}). This case corresponds to the point labeled as ``Scenario $1$'' in Fig.~\ref{Fig6}c, where the storage domain length differential is $\Delta \ell_s =\lambda/4$, achieved with a time delay of $\Delta t =(\ell_{s1}+\ell_{s2})/c$. At this point, the transfer function reaches its minimum value, $H(\omega_t) \approx 0.08$, yielding destructive interference and, consequently, near-complete suppression of the target frequency component. This confirms that a quarter-wavelength path mismatch is sufficient to cancel the target frequency at the output. Alternatively, the inverse outcome ``Scenario $1$ (Inv)'' is obtained by shifting the operating point along the same curve to $\Delta \ell_s =\lambda/2$, corresponding to the same time delay of $\Delta t =(\ell_{s1}+\ell_{s2})/c$. In this case, the two wave paths recombine at the same phase, driving the transfer function to its maximum value of $H(\omega_t) \approx 2.2$. This constructive interference doubles the amplitude of the target frequency component relative to the reference, $\tilde{u}_1(x_o,y)$.  

To quantify the performance of both cases, Scenario $1$ and Scenario $1$ (Inv), the normalized amplitude change, 
\begin{equation}
\Delta\hat{u}(\omega) = \frac{|\mathcal{F}\{\tilde{u}_c(x_c,y)\}| - |\mathcal{F}\{\tilde{u}_1(x_o,y)\}|}{\max(|\mathcal{F}\{\tilde{u}_1(x_o,y)\}|)},
\label{eq:amp_change}
\end{equation}
is evaluated across the excitation spectrum in Fig.~\ref{Fig6}d. As anticipated, Scenario $1$ (Inv) yields an amplification peak of $\Delta\hat{u}(\omega_t) \approx 1.15$ at the target frequency of $92$ kHz, whereas Scenario $1$ reveals a significant dip of $\Delta\hat{u}(\omega_t) \approx -0.87$ at the same frequency. This design allows us to selectively suppress or amplify a specific spectral component purely through controlled phase manipulation, moving continuously between $\Delta \ell_s =\lambda/4$ and $\lambda/2$ along the same transfer function curve. Together, this input shaper manifests another fundamental delay-enabled operation, highlighting the potential of wave-based mechanical systems for advanced physical signal processing.

\begin{figure*}[ht!]
\centering
\includegraphics[width=\linewidth]{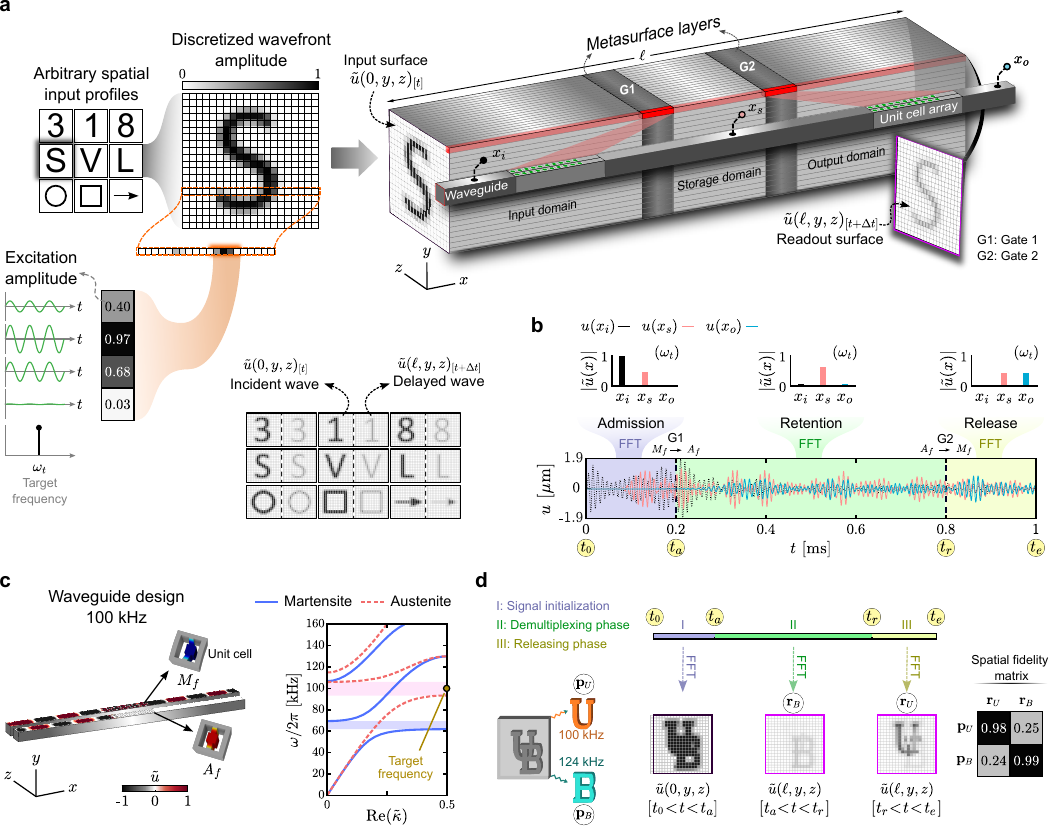}
\caption{\textbf{Mechanical image recovery and decomposition in three-dimensional waveguides.} (a) Conceptual illustration of the system applying a time delay, $\Delta t$, to the incident spatial profile, $\tilde{u}(0,y,z)_{[t]}$. The left panels depict the preprocessing and spatial discretization of an arbitrary input profile (`S' as an example) before being introduced to the waveguide. The right panel demonstrates the propagation of the encoded signal through the system and its delayed readout, $\tilde{u}(\ell,y,z)_{[t+\Delta t]}$. The bottom matrix compares the incident spatial patterns with the corresponding delayed readouts, confirming the preservation of spatial information after a controlled delay. (b) Normalized displacement amplitude at three probes, $x_i$, $x_s$, and $x_o$, placed within the input, storage, and output domains, respectively, alongside their frequency-domain magnitudes at a target frequency of $100$ kHz. (c) Waveguide design and corresponding dispersion diagrams for the image processing system, illustrating the underlying bandgap regions. (d) Demonstration of image separation using an interlocking `UB' logo via frequency demultiplexing. The timeline and associated readout surfaces depict the image distillation process, in which the `B' is recovered and released over the interval $[t_a, t_r]$ at $124$ kHz, while the `U' is preserved and subsequently released over the interval $[t_r, t_e]$ at $100$ kHz. The spatial-pattern MAC matrix at the far right quantifies the similarity between prescribed input patterns (${\mathbf{p}_U}$, ${\mathbf{p}_B}$; rows) and recovered output patterns (${\mathbf{r}_U}$, ${\mathbf{r}_B}$; columns).}
\label{Fig7}
\end{figure*}

\section{Three-dimensional memory and acousto-mechanical image distillation}
\label{3D}

In this final section, we combine the different attributes shown thus far into a three-dimensional (3D) system, in an attempt to showcase higher-level signal manipulation capabilities in a practical, geometrically involved setting. The objective is to depict the system's capacity to process tangible, multi-dimensional spatiotemporal information in mechanical form, demonstrating an ability to identify, store, and break down data (e.g., acoustic-based images) into constitutive elements. We start by extruding the locally resonant unit cell by $20$ mm (Fig.~\ref{Fig7}a) and adjusting the target frequency to $100$ kHz to achieve the desired performance at the new size. The 3D unit cell consists of a rectangular Nitinol resonator ($a=3$ mm and $b=3.15$ mm) connected via two narrow necks ($t_n=1.59$ mm) to a steel frame. The frame has a total width of $l_t=10.13$ mm and an outer width of $t_o=0.065$ mm, with horizontal and vertical thicknesses of $t_k=1.51$ mm and $t_b=1.22$ mm, respectively. The updated dispersion diagram of the extruded cell, shown in Fig.~\ref{Fig7}c, reveals that $100$ kHz falls within the local resonance bandgap when the SMA resonator resides within the austenite phase, $A_f$, and that impinging waves would freely propagate through the waveguide upon transitioning to the martensite phase, $M_f$. A frequency-domain study is conducted at the target frequency on a waveguide composed of $20$ unit cells arranged in two arrays to evaluate the transmission characteristics (Fig.~\ref{Fig7}c). The waveguide is then expanded into a $20\times20$ array of $400$ parallel channels, forming a 3D structure of total length $\ell$ with two metasurface layers (Gates $1$ and $2$), as illustrated in Fig.~\ref{Fig7}a. In the present model, the $400$ spatially separated channels are treated as mechanically uncoupled waveguides. Each channel is excited by a uniform amplitude signal corresponding to one pixel of the incident spatial pattern, and independently performs the prescribed admission, confinement, and release sequence. This arrangement of channels performs the temporary storage and retrieval of the spatially encoded patterns. Recovering the input pattern depends on both the performance of individual channels as well as the consistency of their release times.

In the 3D setting, $x$ remains the primary information transmission (wave propagation) direction, $y$ remains the vertical coordinate, and $z$ denotes the depth. As such, a deformation $u=\tilde{u}(x,y,z) e^{\mathrm{i} \omega t}$ represents the displacement at any given location for waves oscillating at a frequency $\omega$, and $\tilde{u}$ indicates the spatial component of the wavefield. The platform is tested using a variety of inputs, ranging from alphanumeric characters to geometric shapes. The input port utilized in earlier sections is now replaced with an input surface at $x=0$. As shown in Fig.~\ref{Fig7}, each character or shape is incident on the input surface in the form of an edge displacement, $\tilde{u}(0,y,z)_{[t]}$, at an arbitrary time instant, $t$. The computing system then stores the information in the domain flanked by both gates and releases it after a delay, $\Delta t$, where the output is read out at a downstream location, $x_o$.

In the example shown, the letter `S' is discretized into $20\times20$ grayscale pixels, with the color intensity of each pixel corresponding to the amplitude of a boundary displacement applied at the input surface of the associated waveguide. The resulting $400$ nodal displacements form the image of the letter which is fed as a two-dimensional input wave to the system as a harmonic excitation at the target frequency. As established earlier, the pre-configuration thermal states of gates $1$ and $2$ are $M_f$ and $A_f$, respectively. Once the waves enter the storage domain, gate $1$ is thermally activated causing its resonators to go into information-trapping mode. After a prescribed delay, the trapped waves exit the storage domain upon removal of the thermal load on gate $2$, enabling waves to propagate toward the output terminal, $\tilde{u}(\ell,y,z)_{[t+\Delta t]}$. As shown in the images, the spatial features of the signal are preserved, with some attenuation near the pixel edges due to inevitable losses. Similarly, input waves corresponding to the characters `V', `L', `$3$', `$1$', and `$8$', and the geometric shapes `circle', `square', and `right arrow' are successfully delayed and reconstructed at the readout surface, as illustrated in the bottom panel of Fig.~\ref{Fig7}a. To visualize the data storage process, normalized displacement amplitudes at three probe locations, $x_i$ (input), $x_s$ (storage), and $x_o$ (output), are plotted in Fig.~\ref{Fig7}b. During the signal initialization phase, $t_0 < t < t_a$, the wavefield is dominant at the input domain. The energy becomes confined within the storage domain once gate $1$ undergoes a phase transition at $t_a$, as evidenced by the high frequency-domain magnitude captured in the bar chart at $x_s$. The dominant amplitude then shifts to the output probe, $x_o$, once the image is released. The results confirm the scalability of the proposed mechanical processing platform and its effectiveness in storing and releasing higher-dimensional data.

Finally, we demonstrate the proposed architecture's advanced image processing capabilities by showing its ability to decompose multiband images whose sub-elements are housed within different frequency channels. An image containing the University at Buffalo's primary academic logo is fed to the system \cite{UB_Logo}. The image comprises the letter `U' wrapped around and intertwined with the letter `B' to form an interlocking acronym, with the `U' and `B' encoded within $100$ and $124$ kHz, respectively, of the input signal. The computing substrate disentangles the interlocking letters following a three-step procedure similar to that introduced in Sec.~\ref{demultiplexing}, as shown in the timeline of Fig.~\ref{Fig7}d. Attenuated versions of the two constitutive letters are recovered at two different time intervals, as shown from the FFT of the time-domain signal at the readout surface. The spatial-pattern recovery in Fig.~\ref{Fig7}d is quantified using a Modal Assurance Criterion (MAC)-based similarity matrix to evaluate the separation of the two patterns `U' and `B'. The input reference patterns are extracted from the mixed incident field, $\tilde{u}(0,y,z)_{[t]}$, during the interval $[t_0,t_a]$, using the Fourier amplitudes at $100$ and $124$ kHz, respectively. The recovered `B' pattern is extracted from $\tilde{u}(\ell,y,z)_{[t]}$ at $124$ kHz during $[t_a,t_r]$, while the recovered `U' pattern is extracted at $100$ kHz during $[t_r,t_e]$. Each spatial amplitude map is arranged into a vector of $400$ values corresponding to the channel array. The similarity matrix is defined as,
\begin{equation}
{\mathrm{MAC}}_{\alpha\beta}=\frac{{({\mathbf{p}}_{\alpha}\cdot {\mathbf{r}}_{\beta})}^{2}}{({\mathbf{p}}_{\alpha}\cdot {\mathbf{p}}_{\alpha})({\mathbf{r}}_{\beta}\cdot {\mathbf{r}}_{\beta})},\quad \quad \alpha,\beta\in{\{\mathrm{U},\mathrm{B}\}},
\end{equation}
where ${\mathbf{p}}_{\alpha}$ and ${\mathbf{r}}_{\beta}$ denote the prescribed spatial amplitude vectors of the input reference pattern, $\alpha$, and the recovered output pattern, $\beta$, respectively, evaluated at corresponding locations and time instances across the $20\times20$ channel array. Matrix rows represent input references while columns represent recovered output values. Diagonal values evaluate recovery of the intended spatial pattern, whereas off-diagonal values evaluate similarity to the alternative pattern. The matrix shown in Fig.~\ref{Fig7}d confirms the strong separation of the patterns, with high diagonal values near $1$ indicating successful spatial pattern recovery and limited shape mixing, and low off-diagonal values indicating lower similarity to the alternative input pattern.

\section{Conclusions}
\label{conclusion}

In summary, we have numerically demonstrated advanced mechanical signal processing via programmable wave-based mechanical computing. By dynamically tuning the mechanical impedance of SMA-based metasurfaces, we realize functionalities including signal delay, frequency demultiplexing, and input shaping. The system's primary operational mechanism relies on a set of thermally reconfigurable metasurface gates whose dispersion characteristics switch between passband and bandgap states through the martensite-to-austenite phase transitions of the underlying SMA resonators, enabling selective admission, retention, and release of information residing within elastic waves. The concept was employed to strategically admit, trap, and release data for a different set of end goals. The demonstrated analog delay of spatially complex wave packets verified that the proposed architecture preserves signal integrity during temporal storage and release, and enabled selective extraction of different signal components from broadband inputs. The system also revealed effective temporal synchronization of parallel waves to achieve constructive or destructive interference, and its scalability was further demonstrated by extending the system to three-dimensional waveguides for mechanical image processing and decomposition.

It is important to note that the present results were obtained under idealized mechanical constraints and prescribed SMA state transitions. However, while the baseline simulations neglected material damping, the subsequent sensitivity study quantified its effects on the recovered amplitude and spatial-waveform fidelity. The model does not resolve heat transfer, thermal hysteresis, or the energy consumption associated with switching. Despite these assumptions, reflection, leakage, and scattering can still alter the retained and released amplitudes. Therefore, it is critical during physical implementation that material dissipation and finite heating and cooling times be considered together when determining the achievable storage duration, such that each gate reaches its intended state within the available window for confinement or release. On the flip side, it is also important to emphasize that the metasurface gates do not need to switch at a pace that is commensurate with the high ultrasonic carrier frequency chosen here, but they rather switch only at each storage and release event. Prior experiments have demonstrated SMA-based bandgap tuning \cite{candido2018tunable}, and thermally controlled switching of a related metamaterial wavegate \cite{fort2025thermoelastic}. These studies support the constituent mechanisms, although they do not establish the rapid switching or signal retention demonstrated numerically here. Physical implementation of the present mechanical signal processing system requires fabrication and integration of a large number of resonators, uniform thermal switching of the two gates, and preservation of signal amplitude during confinement. Rapid transformation cycles (as fast as $200$ ns) reported in nanoscale NiTi films suggest that fast switching may be possible under suitable thermal conditions \cite{lunser2024transforming}, although achieving such control across the present resonators remains a practical challenge. While localized Joule or laser heating and engineered cooling paths could facilitate thermal control, rapid switching may require innovative and effective heat removal mechanisms. These considerations motivate further evaluation of physical signal retention under experimentally representative conditions.

More broadly, this work establishes a comprehensive framework for programmable temporal control of elastic waves. Although thermal activation is employed here to reconfigure the metasurfaces, the underlying design principles are independent of the specific triggering mechanism and can be extended to alternative stimuli-responsive metamaterials, broader frequency regimes, and different wave modes. By enabling controllable storage, synchronization, and release of mechanical information directly within the propagation medium, the proposed architecture provides a fundamental building block for next-generation wave-based mechanical computing and opens new opportunities for applications ranging from advanced signal processing to physical reservoir computing and beyond.

\section*{Acknowledgments}
This work was supported by the U.S. Office of Naval Research under award no. N000142612090.

\section*{ORCID iD}
\noindent Saeed Behboodi: \url{https://orcid.org/0000-0003-4129-0908}\\
Mohamed Mousa: \url{https://orcid.org/0000-0002-3561-9407}\\
Mostafa Nouh: \url{https://orcid.org/0000-0002-2135-5391}

\section*{Declaration of conflicting interests}
The authors declare no potential conflicts of interest with respect to the research, authorship, and/or publication of this article.

\section*{Data availability statement}
The data supporting the findings of this study are available from the corresponding author upon reasonable request. All relevant data generated or analyzed during this study have been included in the article.

\bibliographystyle{elsarticle-num}
\bibliography{references}



\end{document}